# Investigation of microwave radiation from a compressed beam of ions using generalized Planck's radiation law

Sreeja Loho Choudhury[1], R. K. Paul[2]

Department of Physics, Birla Institute of Technology, Mesra, Ranchi-835215, Jharkhand, India

**Abstract**

An ion-beam compressed by an external electric force is characterized by a unique non-equilibrium distribution function. This is a special case of Tsallis distribution with entropy index q=2, which allows the system to possess appreciably low thermal energy. The thermal radiation by such compressed ion-beam has been investigated in this work. As the system is non extensive, Planck's law of radiation has been modified using Tsallis thermostatistics for the investigation of the system. The average energy of radiation has been derived by introducing the non extensive partition function in the statistical relation of internal energy. The spectral energy density, spectral radiation and total radiation power have also been computed. It is seen that a microwave radiation will be emitted by the compressed ion-beam. The fusion energy gain Q (ratio of the output fusion power to the power consumed by the system) according to the proposed scheme (R. K. Paul 2015) using compressed ion-beam by electric field will not change significantly as the radiated power is very small.

[1] *Email address:* dreamysreeja@gmail.com

[2] *Email address:* ratan_bit1@rediffmail.com

# 1. Introduction

Earlier, an attempt was made to generalize the Planck's radiation law [1] for the explanation of the cosmic microwave background radiation [2] at a temperature of 2.725 K. There are some versions of generalized Planck's law available in the existing literature [3] in this regard. There are also recent attempts to generalize the Planck's radiation law using Kaniadakis approach [4, 5]. In 2015, distribution function of an ion-beam under compression by an external electric field was derived by R. K. Paul. It is known that an ion-beam under compression by an external electric field is characterized by a unique non-equilibrium distribution [6]. This is a special case of Tsallis distribution [7] with entropy index q=2. The characteristic feature of this distribution is that the system possesses a very small thermal energy [6, 7]. The main objective of this work is to investigate the thermal radiation from such a compressed ion-beam.

The thermal radiation from an object at any equilibrium temperature obeys Planck's radiation law [8]. However, since the system of ion-beam under compression by an external electric field is non extensive, explanation of its thermal radiation demands modification of Planck's law of radiation. Here, we have derived the average energy of radiation by introducing the non extensive partition function [9] in the statistical relation of internal energy. From the average energy, we have computed the spectral energy density and spectral radiance of the emitted radiation. It has been seen that this expression of average energy recovers the original Planck's relation for q= 1 (extensive).

The current paper shows that a microwave radiation will be emitted by the compressed ion-beam. The characteristic feature of this radiation in terms of peak wavelength, peak amplitude of radiance and the total radiation power are also computed.

# 2. Method

Let us consider a cube of side L having conducting walls which is filled with electromagnetic radiation in thermal equilibrium at a temperature T. The radiation emitted from a small hole in one of the walls will be characteristic of a perfect black body.

According to statistical mechanics, the probability distribution over the energy levels of a particular mode is given by:

$$P = \frac{\exp(-\beta E_n)}{Z} \tag{1}$$

where $E_n = \left(n + \frac{1}{2}\right) h\nu$, n=0,1,2,3,..... and partition function

$$Z = \sum_{n=0}^{\infty} \exp(-\beta E_n) \tag{2}$$

Ordinary statistical mechanics is derived by maximizing the Boltzmann Gibbs entropy given by:

$$S = -k \sum_i p_i \ln p_i \tag{3}$$

which is subjected to constraints, whereas in case of non extensive statistical mechanics the more general Tsallis entropy [10] given by:

$$S_q = \frac{k(1 - \sum_i p_i^q)}{(q-1)} \tag{4}$$

is maximized. Here $p_i$ are probabilities associated with the microstates of a physical system and q is the non extensive entropic index. The ordinary Boltzmann Gibbs entropy is obtained in the limit q→1. For some given set of probabilities $p_i$, one can proceed to another set of probabilities $P_i$ given as:

$$P_i = \frac{p_i^q}{\sum_i p_i^q} \tag{5}$$

The probability $P_i$ coming out after maximizing $S_q$ under the energy constraint $\sum_i P_i \epsilon_i = U_q$, where $\epsilon_i$ are the energy levels of the microstates is:

$$P_i = \frac{(1+(q-1)\beta \epsilon_i)^{\frac{-q}{(q-1)}}}{Z_q} \tag{6}$$

where $Z_q = \sum_i (1 + (q-1)\beta \epsilon_i)^{\frac{-q}{(q-1)}}$ is the partition function [7] in case of non extensive statistical mechanics and $\beta = \frac{1}{kT}$, $k$= Boltzmann constant.

If we consider $i$=n then,

$$Z_q = \sum_{n=0}^{\infty} \{1 - (1-q)\beta E_n\}^{\frac{q}{(1-q)}} \tag{7}$$

$$Z_q = \sum_{n=0}^{\infty}\{1 - (1-q)\left(n+\frac{1}{2}\right)\frac{h\nu}{kT}\}^{\frac{q}{(1-q)}} \tag{8}$$

## 2.1 Calculation of Average energy

The average energy in a mode can be expressed in terms of the partition function as:

$$<E> = kT^2 \frac{\partial}{\partial t} \ln Z_q \tag{9}$$

Now

$$\frac{\partial}{\partial t} \ln Z_q = \frac{1}{\sum_{n=0}^{\infty}\{1-(1-q)\left(n+\frac{1}{2}\right)\frac{h\nu}{kT}\}^{\frac{q}{(1-q)}}} \left[\frac{q}{(1-q)}\sum_{n=0}^{\infty}\{1-(1-q)\left(n+\frac{1}{2}\right)\frac{h\nu}{kT}\}(1-q)\left(n+\frac{1}{2}\right)\frac{h\nu}{kT^2}\right]$$

Let

$$\frac{\partial}{\partial t}\ln Z_q = \frac{qh\nu}{kT^2}\frac{a}{b} \tag{10}$$

Now,

$$b = \sum_{n=0}^{\infty}\{1-(1-q)\left(n+\frac{1}{2}\right)\frac{h\nu}{kT}\}^{\frac{q}{(1-q)}} \tag{11}$$

From the definition of limit we have, $\lim_{n\to\infty}\left(1+\frac{x}{n}\right)^n = e^x$.

$$b = \{1-(1-q)\frac{h\nu}{2kT}\}^{\frac{q}{(1-q)}} + \{1-(1-q)\frac{3h\nu}{2kT}\}^{\frac{q}{(1-q)}} + \{1-(1-q)\frac{5h\nu}{2kT}\}^{\frac{q}{(1-q)}} + \cdots \infty \tag{12}$$

as q→1, (1-q)→0, and $\frac{q}{(1-q)} \to \infty$ then

$$b = e^{\frac{-qh\nu}{2kT}} + e^{\frac{-3qh\nu}{2kT}} + e^{\frac{-5qh\nu}{2kT}} + \cdots \infty \tag{13}$$

Therefore,

$$b = \frac{e^{\frac{qh\nu}{2kT}}}{(e^{\frac{qh\nu}{kT}}-1)} \tag{14}$$

From (10) we have,

$$a = \sum_{n=0}^{\infty}\left(n+\frac{1}{2}\right)\{1-(1-q)\left(n+\frac{1}{2}\right)\frac{h\nu}{kT}\}^{\frac{2q-1}{(1-q)}} \tag{15}$$

Let

$$a = c + d$$

where $d = \frac{1}{2}\sum_{n=0}^{\infty}\{1 - (1-q)(n+\frac{1}{2})\frac{h\nu}{kT}\}^{\frac{2q-1}{(1-q)}}$ (16)

$$d = \frac{1}{2}[\{1 - (1-q)\frac{h\nu}{2kT}\}^{\frac{2q-1}{(1-q)}} + \{1 - (1-q)\frac{3h\nu}{2kT}\}^{\frac{2q-1}{(1-q)}} + \{1 - (1-q)\frac{5h\nu}{2kT}\}^{\frac{2q-1}{(1-q)}} + \cdots \infty]$$

Therefore,

$$d = \frac{1}{2}[\frac{e^{(2q-1)\frac{h\nu}{2kT}}}{\{e^{(2q-1)\frac{h\nu}{kT}}-1\}}]$$ (17)

Now,

$$c = \sum_{n=0}^{\infty} n\{1 - (1-q)(n+\frac{1}{2})\frac{h\nu}{kT}\}^{\frac{2q-1}{(1-q)}}$$ (18)

$$c = \frac{e^{(2q-1)\frac{h\nu}{2kT}}}{\{e^{(2q-1)\frac{h\nu}{kT}}-1\}^2}$$ (19)

Hence,

$$a = \frac{e^{(2q-1)\frac{h\nu}{2kT}}}{\left(e^{(2q-1)\frac{h\nu}{kT}}-1\right)}\left(\frac{1}{2} + \frac{1}{e^{(2q-1)\frac{h\nu}{kT}}-1}\right)$$ (20)

Therefore,

$$\frac{a}{b} = \frac{(e^{(2q-1)\frac{h\nu}{kT}}-e^{(q-1)\frac{h\nu}{kT}})}{\left(e^{(2q-1)\frac{h\nu}{kT}}-1\right)}\left(\frac{1}{2} + \frac{1}{e^{(2q-1)\frac{h\nu}{kT}}-1}\right)$$ (21)

From (9) we have the average energy given as:

$$<E> = \frac{(e^{(2q-1)\frac{h\nu}{kT}} - e^{(q-1)\frac{h\nu}{kT}})}{\left(e^{(2q-1)\frac{h\nu}{kT}} - 1\right)}\left(\frac{1}{2} + \frac{1}{e^{(2q-1)\frac{h\nu}{kT}} - 1}\right)$$

For q=1 (extensive), we have

$$<E> = \frac{h\nu}{2} + \frac{h\nu}{(e^{\frac{h\nu}{kT}}-1)}$$ (22)

Neglecting the vacuum energy term $\frac{h\nu}{2}$,

$$<E> = \frac{h\nu}{(e^{\frac{h\nu}{kT}}-1)} \qquad (23)$$

For q=2,

$$<E> = 2h\nu \frac{(e^{\frac{3h\nu}{kT}}-e^{\frac{h\nu}{kT}})}{\left(e^{\frac{3h\nu}{kT}}-1\right)} \left(\frac{1}{2} + \frac{1}{e^{\frac{3h\nu}{kT}}-1}\right) \qquad (24)$$

Neglecting the vacuum term,

$$<E> = 2h\nu \frac{(e^{\frac{3h\nu}{kT}}-e^{\frac{h\nu}{kT}})}{\left(e^{\frac{3h\nu}{kT}}-1\right)^2} \qquad (25)$$

Here $h\nu \gg kT$ and hence we have,

$$<E> = 2h\nu(e^{\frac{3h\nu}{kT}} - e^{\frac{h\nu}{kT}})e^{\frac{-6h\nu}{kT}} \qquad (26)$$

$$<E> = 2h\nu(e^{\frac{-3h\nu}{kT}} - e^{\frac{-5h\nu}{kT}}) \qquad (27)$$

Similarly, we can have the average energy for other q values such as q=0.95, q=1.5, etc.

## 2.2 Energy Density

The energy density per unit frequency is given by:

$$u_\nu(T) = \frac{8\pi\nu^2}{c^3}<E> \qquad (28)$$

Therefore,

$$u_\nu(T) = \frac{8\pi h\nu^3 q}{c^3} \frac{(e^{(2q-1)\frac{h\nu}{kT}}-e^{(q-1)\frac{h\nu}{kT}})}{(e^{(2q-1)\frac{h\nu}{kT}}-1)^2} \qquad (29)$$

Hence the generalized energy density per unit frequency is given by:

$$u_\nu(T) = \frac{8\pi h\nu^3 q}{c^3} \frac{\{e^{(2q-1)\frac{h\nu}{kT}} - e^{(q-1)\frac{h\nu}{kT}}\}}{\left\{e^{(2q-1)\frac{h\nu}{kT}} - 1\right\}^2}$$

For q=1,

$$u_\nu(T) = \frac{8\pi h\nu^3}{c^3(e^{\frac{h\nu}{kT}}-1)} \tag{30}$$

For q=0.95,

$$u_\nu(T) = \frac{7.6\pi h\nu^3}{c^3} \frac{\left(e^{\frac{0.9h\nu}{kT}} - e^{\frac{-0.05h\nu}{kT}}\right)}{\left(e^{\frac{0.9h\nu}{kT}}-1\right)^2} \tag{31}$$

For q=1.5,

$$u_\nu(T) = \frac{12\pi h\nu^3}{c^3}\left(e^{\frac{-2h\nu}{kT}} - e^{\frac{-3.5h\nu}{kT}}\right) \tag{32}$$

Similarly for q=2,

$$u_\nu(T) = \frac{16\pi h\nu^3}{c^3}\left(e^{\frac{-3h\nu}{kT}} - e^{\frac{-5h\nu}{kT}}\right) \tag{33}$$

The variation of energy density with frequency for different q values has been shown in the following figure 1 for the temperature T=2.725K.

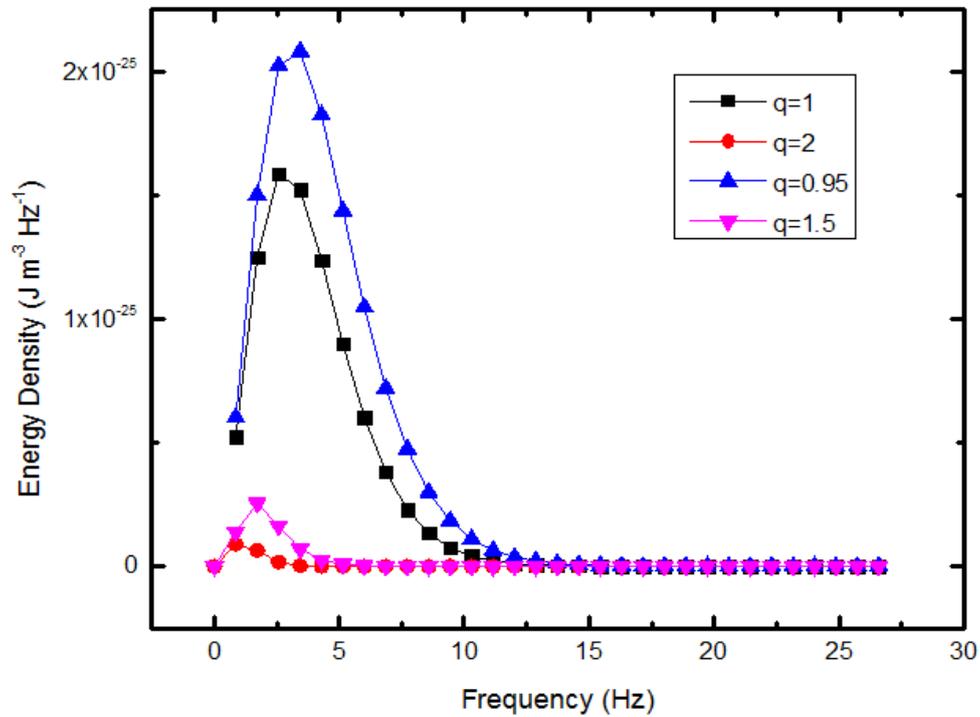

FIGURE 1. The plot of energy density versus frequency for different q values at a temperature T= 2.725 K.

## 2.3 Spectral Radiance

The spectral radiance is given by:

$$I_\nu(T) = \frac{u_\nu(T)c}{4} \tag{34}$$

We have, frequency of radiation $\nu = \frac{c}{\lambda}$

Therefore, $\frac{d\nu}{d\lambda} = \frac{-c}{\lambda^2}$

So we get,

$$I_\lambda(T) = I_\nu(T)\left|\frac{d\nu}{d\lambda}\right| \tag{35}$$

$$I_\lambda(T) = \frac{2\pi hc^2 q}{\lambda^5} \frac{\left\{e^{(2q-1)\frac{hc}{\lambda kT}} - e^{(q-1)\frac{hc}{\lambda kT}}\right\}}{\left\{e^{(2q-1)\frac{hc}{\lambda kT}} - 1\right\}^2} \tag{36}$$

Therfore, for q=1 we have,

$$I_\lambda(T) = \frac{2\pi hc^2}{\lambda^5(e^{\frac{hc}{\lambda kT}} - 1)} \tag{37}$$

For q=2,

$$I_\lambda(T) = \frac{4\pi hc^2}{\lambda^5}\left(e^{\frac{-3hc}{\lambda kT}} - e^{\frac{-5hc}{\lambda kT}}\right) \tag{38}$$

Similarly for q=1.5,

$$I_\lambda(T) = \frac{3\pi hc^2}{\lambda^5}\left(e^{\frac{-2hc}{\lambda kT}} - e^{\frac{-3.5hc}{\lambda kT}}\right) \tag{39}$$

For q=0.95,

$$I_\lambda(T) = \frac{1.9\pi hc^2}{\lambda^5} \frac{\left(e^{\frac{0.9hc}{\lambda kT}} - e^{\frac{-0.05hc}{\lambda kT}}\right)}{(e^{\frac{0.9hc}{\lambda kT}} - 1)^2} \tag{40}$$

Figure 2 is the plot of spectral radiance versus wavelength for a particular temperature say T=2.725K. It is clear that the peak amplitude gradually decreases as q is increased. The shift of peak wavelength on the right hand side (higher wavelengths) with the increase in q values is also observed.

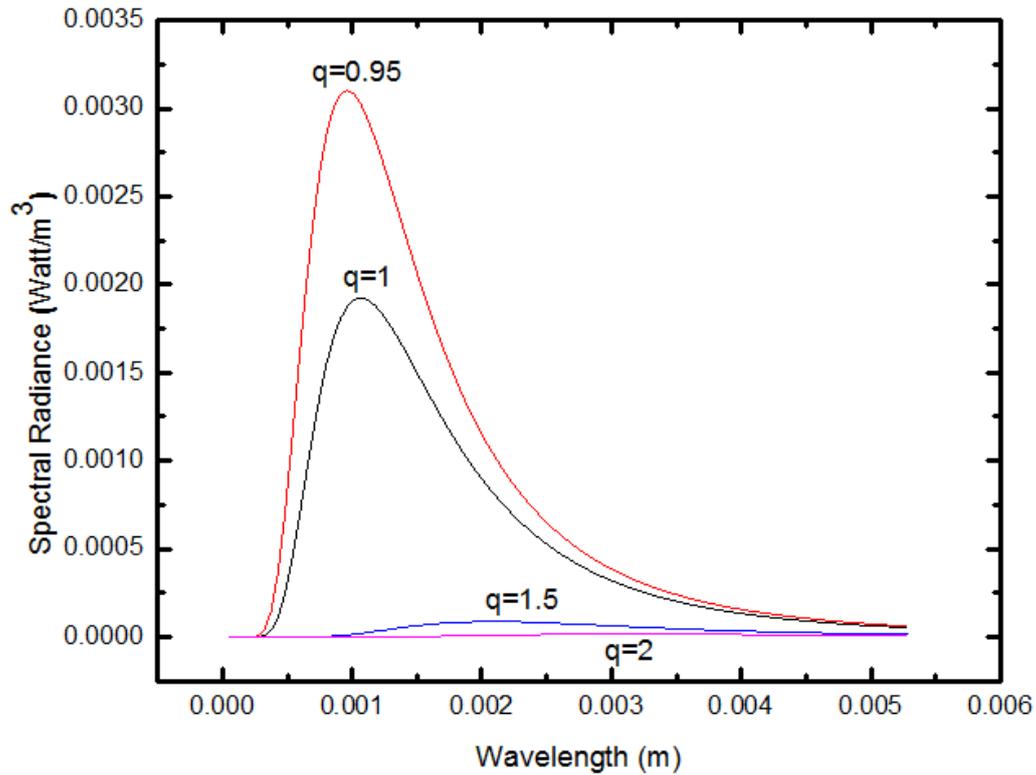

FIGURE 2. The plot of spectral radiance versus wavelength for the entropy index q= 0.95, 1, 1.5 and 2 at a temperature T= 2.725K.

## 3. Results and Discussion

The magnetic confinement approach [11], which is being undertaken worldwide using Tokamak, for the generation of clean and efficient fusion power is based on fusion in plasmas at thermodynamic equilibrium. However, electrostatic confinement approach [12] has also been examined for various plasma systems which are out of thermodynamic equilibrium in terms of their feasibility to produce fusion energy. In both plasma systems, the ion distribution function plays a significant role in obtaining an appropriate value of reaction rate. The characteristic of equilibrium plasma is the Maxwellian distribution, while non-thermodynamic plasmas do not follow this Maxwellian distribution.

An ion-beam under compression by an electrostatic field is characterized by a unique non-equilibrium distribution [6]. The distribution function for a compressed beam from [6] is given as:

$$n_r = N \frac{E_{av}}{\epsilon_c} \left( \frac{e^{(\frac{\epsilon_c}{E_{av}})} - 1}{\epsilon_c} \right) \left\{ \frac{1}{1 + \frac{\left(e^{(\frac{\epsilon_c}{E_{av}})} - 1\right)\epsilon_r}{\epsilon_c}} \right\} \qquad (41)$$

Here N= particle number, $E_{av}$= beam energy (U/N), U= total energy of the system and $\epsilon_c$= upper cut-off energy.

The relation between the thermal energy $\frac{1}{\beta}$ and the beam energy of the system for this non-Maxwellian distribution function [6] is given by:

$$\frac{\frac{1}{\beta}}{E_{av}} = \frac{\text{Thermal energy of each particle}}{\text{Directed energy of each particle}} = 7.45 \times 10^{-11} \qquad (42)$$

From the relation $\beta = \frac{1}{kT}$, temperatures are calculated for different beam energies like 40, 60, 80, 100, 120 keV.

Figure 3 shows the variation of T with $E_{av}$.

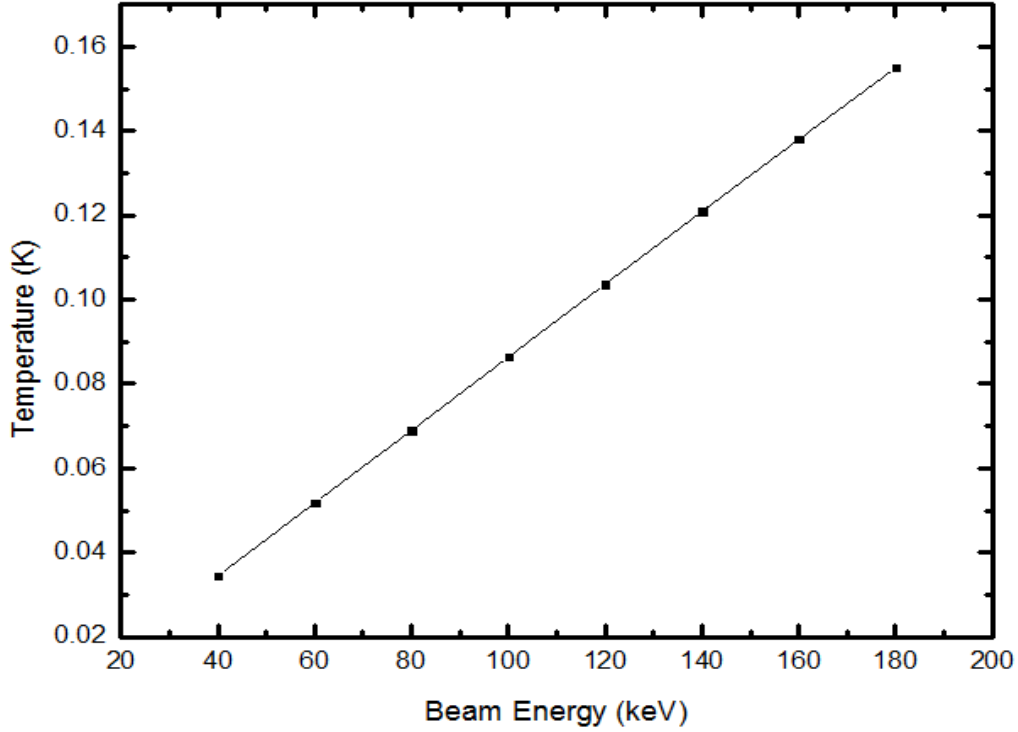

FIGURE 3. The plot of T versus beam energy $E_{av}$.

Temperature of the system increases as the beam energy increases i.e. the relation is linear.

This ion distribution is a special case of Tsallis distribution with the non extensive entropy index q=2. For this distribution the q value of the generalized Planck's radiation formula can be estimated in the following manner.

The ion-beam is subjected to a conservative force, such as the electric field so that the total energy of the system remains constant in the presence of external field change. The temperature of the system can be computed by equating the electrostatic energy $\frac{1}{2}\epsilon_0 E^2$ with the thermal energy $\frac{3}{2}NkT$ for a given particle density N and k is the Boltzmann constant. Therefore,

$$\frac{\frac{1}{2}\epsilon_0 E^2}{N} = \frac{3}{2}kT$$

For particle density N=$10^{20}$ m$^{-3}$ and electric field $E = 4 \times 10^3$ V/m, the temperature of the system comes out to be 0.0342 K. Figure 4 represents the variation of particle temperature as a function of the compressing electrostatic field.

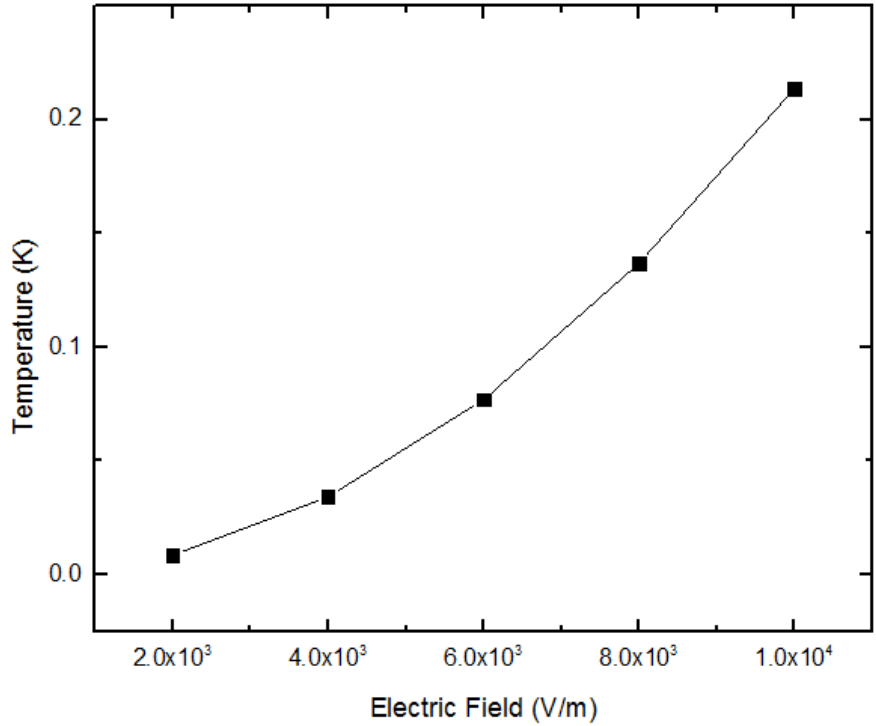

FIGURE 4. Temperature of the system versus compressing electric field

Now, the thermal energy is responsible for the thermal radiation from the compressed beam of ions. Therefore, the energy of radiation is equal to the thermal energy of the system. Further, since thermal energy is equal to the electrostatic energy, it's obvious that the radiation energy equals the electrostatic energy.

$$\frac{\frac{1}{2}\epsilon_0 E^2}{N} = h\nu$$

Therefore,

$$\frac{\frac{1}{2}\epsilon_0 E^2}{N} = h\frac{c}{\lambda}$$

Here c is the speed of light in vacuum, h is the Planck's constant. Substituting the above mentioned values of electric field and particle density, the value of wavelength of thermal radiation from the system comes out to be 0.28 m. Figure 5 represents the variation of wavelength with the changing electric field.

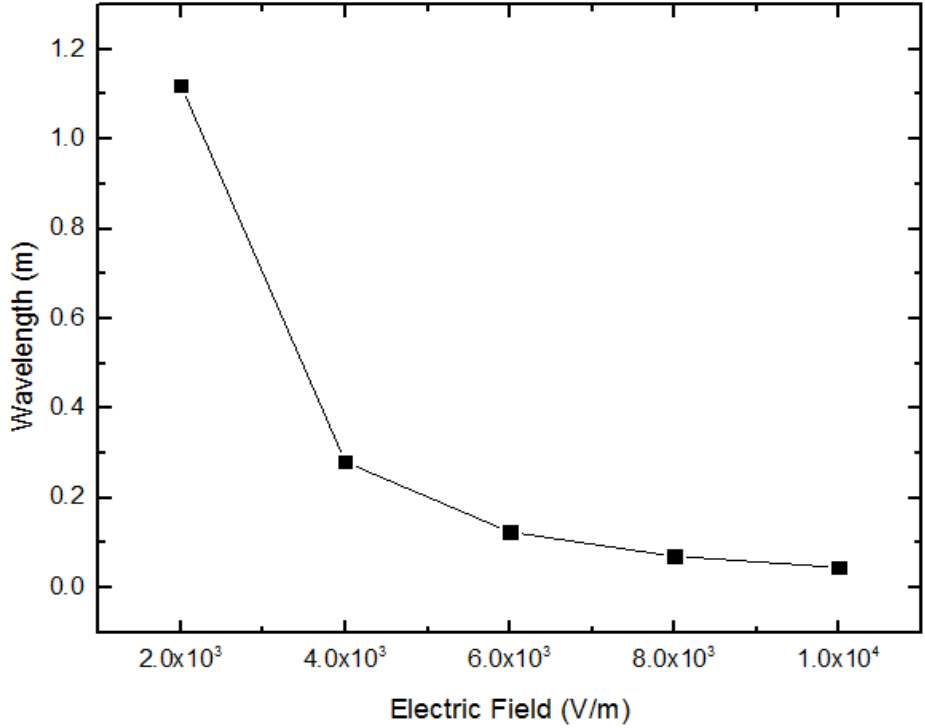

FIGURE 5. Wavelength of thermal radiation versus compressing electric field.

Multiplying the wavelength and temperature we get, $\lambda T = 9.576 \times 10^{-3}\ mK$.

From equation (36), the generalized spectral radiance is given as:

$$I_\lambda(T) = \frac{2\pi h c^2 q}{\lambda^5}\left\{e^{(1-2q)\frac{hc}{\lambda kT}} - e^{(1-3q)\frac{hc}{\lambda kT}}\right\} \tag{43}$$

In order to show that the q value of the generalized Planck's radiation formula is same as the q value of the ion distribution which is a special case of Tsallis distribution with the non extensive entropic index q=2, we have to equate the derivative of $I_\lambda(T)$ with respect to $\lambda$ to zero.

$$\frac{dI_\lambda(T)}{d\lambda} = 0 \tag{44}$$

The equation becomes:

$$e^{(1-2q)\frac{0.0144}{\lambda T}}\left\{5 + (1-2q)\frac{0.0144}{\lambda T}\right\} - e^{(1-3q)\frac{0.0144}{\lambda T}}\left\{5 + (1-3q)\frac{0.0144}{\lambda T}\right\} = 0 \tag{45}$$

Substituting the value of $\lambda T$ in equation (45) we get,

$$q = 2.208$$

i.e,

$$q \approx 2$$

Hence, it is reasonable to take q=2 for generalized Planck's radiation formula corresponding to this ion distribution emanating the thermal radiation.

Therefore, thermal radiation by such compressed ion-beam has been investigated using the expression for spectral radiance for q=2.

$$I_\lambda(T) = \frac{4\pi hc^2}{\lambda^5}\left(e^{\frac{-3hc}{\lambda kT}} - e^{\frac{-5hc}{\lambda kT}}\right)$$

The variation of spectral radiance for different temperatures is shown in figure 6.

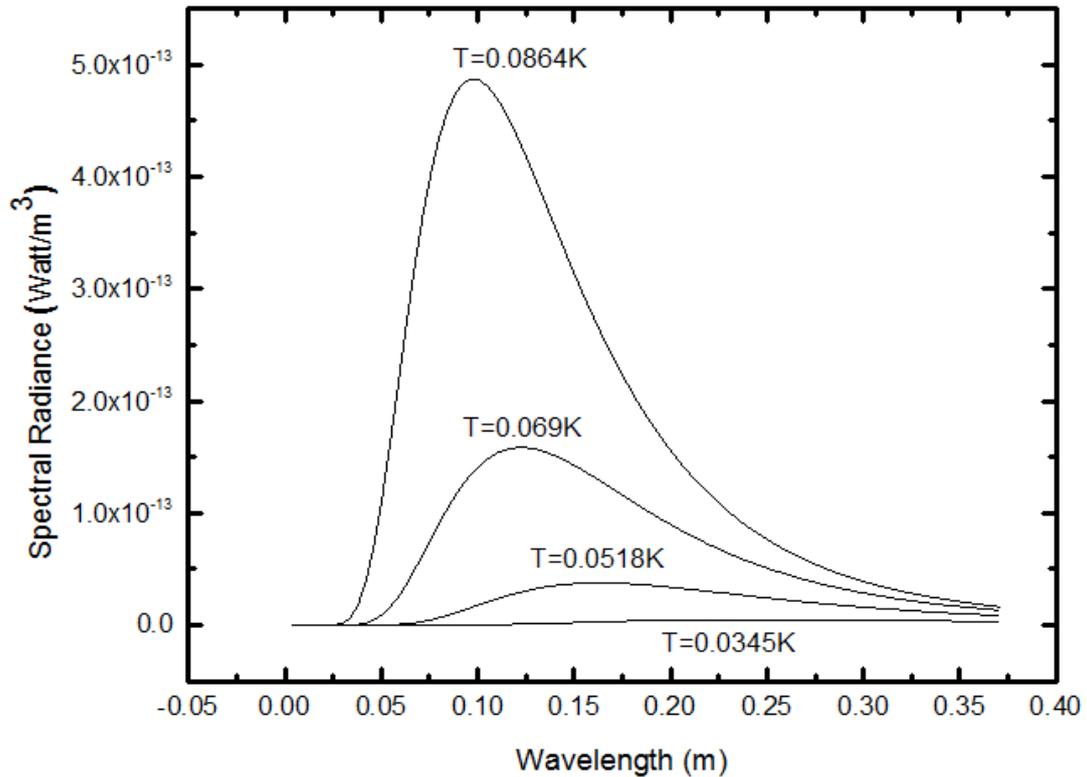

FIGURE 6. The plot of spectral radiance versus wavelength for temperatures T=0.0345, 0.0518, 0.069, 0.0864 K.

The temperatures 0.0345, 0.0518, 0.069, 0.0864 K correspond to the beam energies $E_{av}$= 40, 60, 80, 100 keV respectively. The radiation falls in the microwave radiation range. The peak wavelength for each of the plots can be obtained by making the derivative of $I_\lambda(T)$ with respect to λ equal to zero.

Therefore, $\frac{dI}{d\lambda} = 0$

Let $\frac{hc}{\lambda kT} = x$ then,

$$e^{-3x}(3x - 5) + 5e^{-5x}(1 - x) = 0$$

The solution of this equation is $x=1.705$.

Now, $\lambda_{max}T = \frac{hc}{kx}$ then

$$\lambda_{max}T = 8.448 \times 10^{-3} mK \qquad (46)$$

The peak wavelengths for the temperatures 0.0345, 0.0518, 0.069, 0.0864 K are computed to be 0.2448, 0.163, 0.1224 and 0.097 m respectively. The variation of peak wavelengths with the beam energies has also been plotted in figure 7.

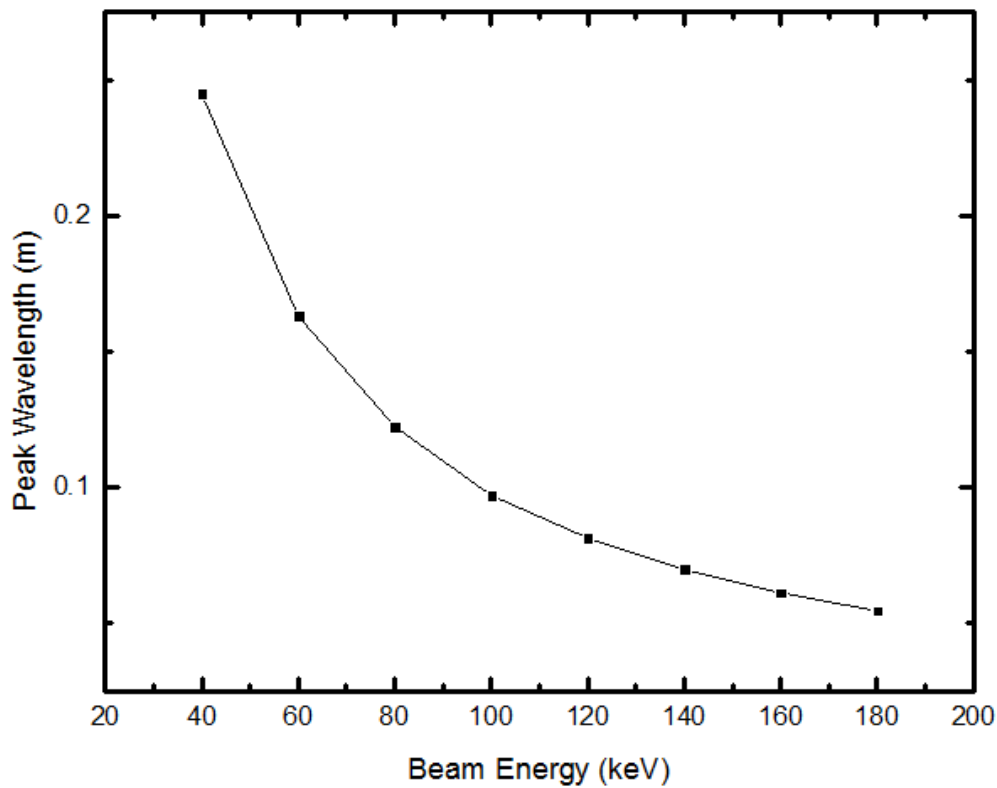

FIGURE 7. The plot of peak wavelength versus beam energy $E_{av}$ (in keV).

The maxima (peak amplitude) of the spectral radiance has also been computed. Figure 8 shows the variation of the peak amplitude with the different beam energies.

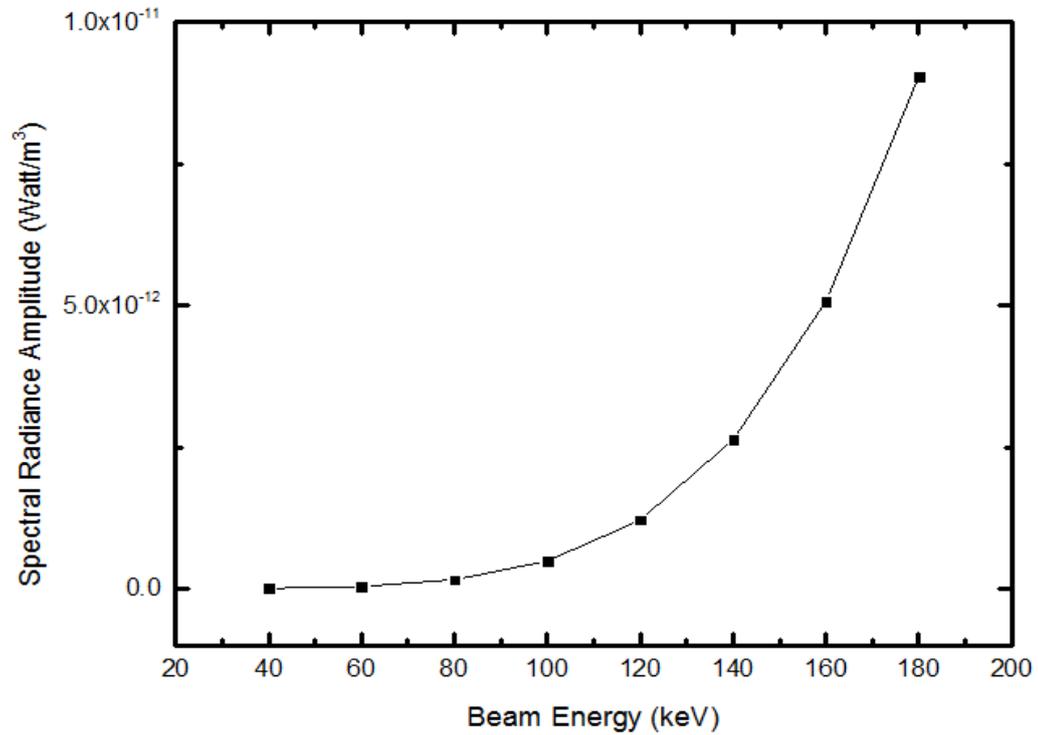

FIGURE 8. The plot of peak amplitude of spectral radiance versus beam energy (in keV).

The total spectral intensity (in Watt/m$^2$) is calculated by integrating the spectral radiance with respect to $d\lambda$ (figure 6). This total radiation intensity is very low of the order $10^{-15}$ to $10^{-13}$. It is possible to detect such a low microwave radiation power [13]. The nature of the radiation intensity is depicted against the beam energy (figure 9).

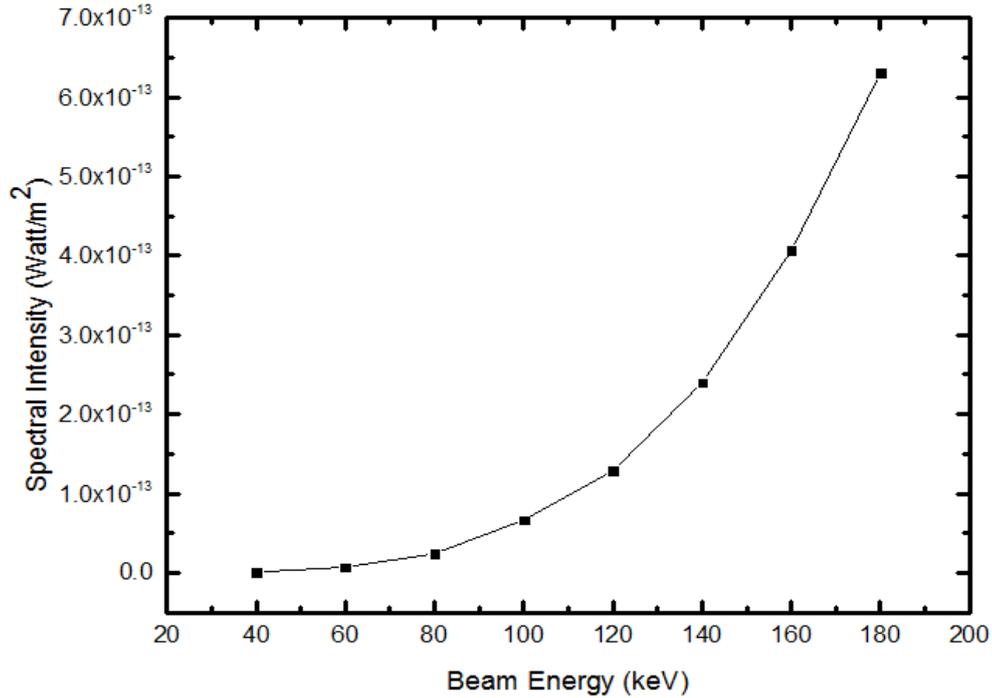

FIGURE 9. The plot of spectral intensity versus beam energy (in keV).

It has been observed that the spectral intensity increases with the increase in beam energy.

# 4. Conclusion

A new method of finding the average energy of a black body is thus derived using the non extensive statistical mechanics formalism. The spectral density and spectral radiance has also been derived following the average energy for entropy index q= 0.95, 1, 1.5, 2. For q=1, the expression of average energy recovers the original form of average energy in case of standard statistical mechanics (extensive). The expression of spectral radiance which has been derived using the proposed method for q=2 is used to investigate the thermal radiation by the compressed ion-beam. It has been observed that a microwave radiation will be emitted from the compressed ion-beam. Since the radiated power is very small of the order of $10^{-15}$, the fusion energy gain Q of the proposed scheme [6] using compressed ion-beam by electric field will not change significantly. However, such microwave radiation will be the precursor to the occurrence of a unique non-equilibrium distribution for the compressed beam of ions.

*Acknowledgements*

The authors are grateful to the Physics department of Birla Institute of Technology, Mesra, Ranchi for rendering an outstanding environment to carry out the research work. The authors also acknowledge the help and support from M. K. Sinha (Physics Department, Birla Institute of Technology, Mesra, Ranchi).